\documentclass[conference]{IEEEtran}

\ifCLASSINFOpdf
\else
\fi
\usepackage{amsmath}
\usepackage{amsfonts}
\usepackage[caption=false,font=small,labelfont=sf,textfont=sf]{subfig}
\usepackage[linesnumbered,ruled,vlined]{algorithm2e}
\usepackage{amssymb}
\usepackage{xcolor}
\usepackage{flushend}
\usepackage{comment}
\usepackage{graphicx}
\usepackage{algorithm2e}
\RestyleAlgo{ruled}
\usepackage{algpseudocode}
\usepackage{amsmath}
\usepackage{textcomp}
\usepackage{epsfig}
\usepackage{multicol}
\usepackage{caption}
\usepackage{algpseudocode}
\usepackage{tikz}
\usepackage{geometry}
\usepackage{pgfplotstable}
    \pgfplotsset{
        compat=1.9,
        compat/bar nodes=1.8,
}

\usepackage{acronym}
\acrodef{ce}[CE]{Channel Estimation}
\acrodef{nr}[NR]{New Radio}
\acrodef{graphnet}[GraphNet]{Graph Neural Estimation Network}
\acrodef{mimo}[MIMO]{Multiple Input Multiple Output}
\acrodef{siso}[SISO]{Single Input Single Output}
\acrodef{ls}[LS]{Least Squares}
\acrodef{mmse}[MMSE]{Minimum Mean Square Error}
\acrodef{mse}[MSE]{Mean Square Error}
\acrodef{dl}[DL]{Deep Learning}
\acrodef{cnn}[CNN]{Convolutional Neural Network}
\acrodef{dmrs}[DM-RS]{Demodulation Reference Signals}
\acrodef{ofdm}[OFDM]{Orthogonal Frequency Division Multiplexing}
\acrodef{mmimo}[mMIMO]{Massive MIMO}
\acrodef{gnn}[GNN]{Graph Neural Network}
\acrodef{re}[RE]{Resource Element}
\acrodef{qam}[QAM]{Quadrature Amplitude Modulation}
\acrodef{awgn}[AWGN]{Additive White Gaussian Noise}
\acrodef{prb}[PRB]{Physical Resource Block}
\acrodef{ut}[UT]{User Terminal}
\acrodef{sr}[SR]{Super Resolution}
\acrodef{ir}[IR]{Image Restoration}
\acrodef{srcnn}[SRCNN]{CNN-based SR}
\acrodef{dncnn}[DnCNN]{CNN-based Denoising}
\acrodef{relu}[ReLU]{Rectified Linear Unit}
\acrodef{nn}[NN]{Neural Network}
\acrodef{dnn}[DNN]{Deep Neural Network}
\acrodef{gcn}[GCN]{Graph Convolutional Network}
\acrodef{cdf}[CDF]{Cumulative Distribution Function}
\acrodef{snr}[SNR]{Signal to Noise Ratio}
\acrodef{bs}[BS]{Base Station}
\acrodef{tdl}[TDL]{Tapped Delay Line}
\acrodef{li}[LI]{Linear Interpolation}
\acrodef{mlp}[MLP]{Multi Layer Perceptron}
\acrodef{pdsch}[PDSCH]{Physical Downlink Shared Channel}
\acrodef{tdl}[TDL]{Tapped Delay Line}
\acrodef{bler}[BLER]{Block Error Rate}
\acrodef{cir}[CIR]{Channel Impulse Response}
\acrodef{csi}[CSI]{Channel State Information}
\begin{document}

\title{CSI Compression Beyond Latents: End-to-End Hybrid Attention-CNN Networks with Entropy Regularization\\
}

\author{
\IEEEauthorblockN{Maryam Ansarifard\IEEEauthorrefmark{1}, Mostafa Rahmani\IEEEauthorrefmark{2}, Mohit K. Sharma\IEEEauthorrefmark{3}, Kishor C. Joshi\IEEEauthorrefmark{1}, George Exarchakos\IEEEauthorrefmark{1}, Alister Burr\IEEEauthorrefmark{2}}

\IEEEauthorblockA{\IEEEauthorrefmark{1}Department of Electrical Engineering, Eindhoven University of Technology, Eindhoven, The Netherlands\\
Emails: \{m.ansarifard, k.c.joshi, G.Exarchakos\}@tue.nl}

\IEEEauthorblockA{\IEEEauthorrefmark{2}School of Physics, Engineering, and Technology, University of York, York, YO10 5DD, UK\\
Emails: \{rahmani.mostafa, alister.burr\}@york.ac.uk}

\IEEEauthorblockA{\IEEEauthorrefmark{3}Directed Energy Research Center, Technology Innovation Institute, Abu Dhabi, UAE\\
Email: mohit.sharma@tii.ae}
}

%\author{\IEEEauthorblockN{1\textsuperscript{st} Maryam Ansarifard}
%\and

%}

\maketitle

\begin{abstract}
Massive MIMO systems rely on accurate Channel State Information (CSI) feedback to enable high-gain beamforming. However, the feedback overhead scales linearly with the number of antennas, presenting a major bottleneck. While recent deep learning methods have improved CSI compression, most overlook the impact of quantization and entropy coding, limiting their practical deployability. In this work, we propose an end-to-end CSI compression framework that integrates a Spatial Correlation-Guided Attention Mechanism with quantization and entropy-aware training. Our model effectively exploits the spatial correlation among the antennas, thereby learning compact, entropy-optimized latent representations for efficient coding. This reduces the required feedback bitrates without sacrificing reconstruction accuracy, thereby yielding
a superior rate-distortion trade-off. Experiments show that our method surpasses existing end-to-end CSI compression schemes, exceeding benchmark performance by an average of $21.5 \%$ on indoor datasets and $18.9 \%$ on outdoor datasets. The proposed framework results in a practical and efficient CSI feedback
scheme.
\end{abstract}


\section{Introduction}
Massive Multiple-Input Multiple-Output (MIMO) systems have emerged as a cornerstone of next-generation wireless communication networks, offering significant gains in spectral and energy efficiency. A critical enabler of these gains is the availability of accurate Channel State Information (CSI) at the base station (gNB). However, in Frequency Division Duplex (FDD) massive MIMO systems, obtaining CSI feedback from the receiver to the transmitter introduces a prohibitive overhead which scales linearly with the number of antennas, thus posing a major bottleneck for practical deployments.\\
\indent To address the CSI feedback problem, various approaches have been explored. Classical Compressive Sensing (CS) methods leverage the sparsity of the wireless channel to reduce feedback requirements \cite{kuo2012compressive} but often struggle under practical channel conditions and require complex recovery algorithms. Deep Learning (DL) techniques, particularly autoencoders, have shown significant promise in learning compact CSI representations. CsiNet \cite{csinet} pioneered Convolutional Neural Network (CNN)-based CSI compression using an Auto-Encoder (AE), with an encoder at the UE and a decoder at the gNB. However, CNN performance heavily depends on the size of the receptive field (or convolutional kernel). To address this, CsiNet+ \cite{csinet+} enhances the original CsiNet by enlarging the receptive field, better exploiting CSI sparsity in the angular-delay domain, while small kernels capture fine details more effectively. Based on this, CRNet \cite{crnet} introduces a multi-resolution design, using different kernel sizes in both the encoder and decoder, enabling more adaptive and effective CSI feedback. Nevertheless, they still perform poorly in outdoor scenarios with high compression ratios (CRs), a common issue for existing algorithms. To address this, CsiNet+DNN \cite{csinet+dnn} introduces additional layers and incorporates a different activation function to acquire more knowledge to enhance the performance of the RefineNet.\\
\begin{figure*}
    \centering
    \subfloat[\footnotesize{UE side includes CNN-transformer based encoder, quantization, and entropy encoding}\label{encoder}]{
        \includegraphics[width=0.7\linewidth]{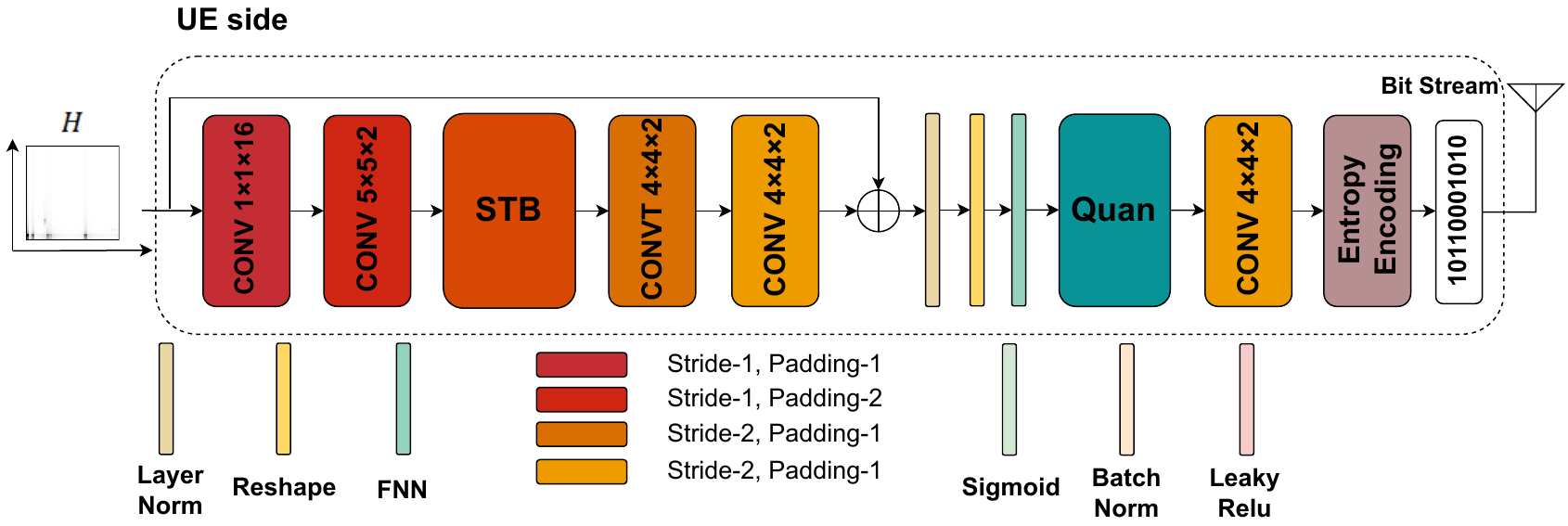}
    }   
    \vspace{1em}   
    \subfloat[\footnotesize{gNB side includes entropy decoder, dequantization, transformer-CNN based decoder}\label{decoder}]{
        \includegraphics[width=0.7\linewidth]{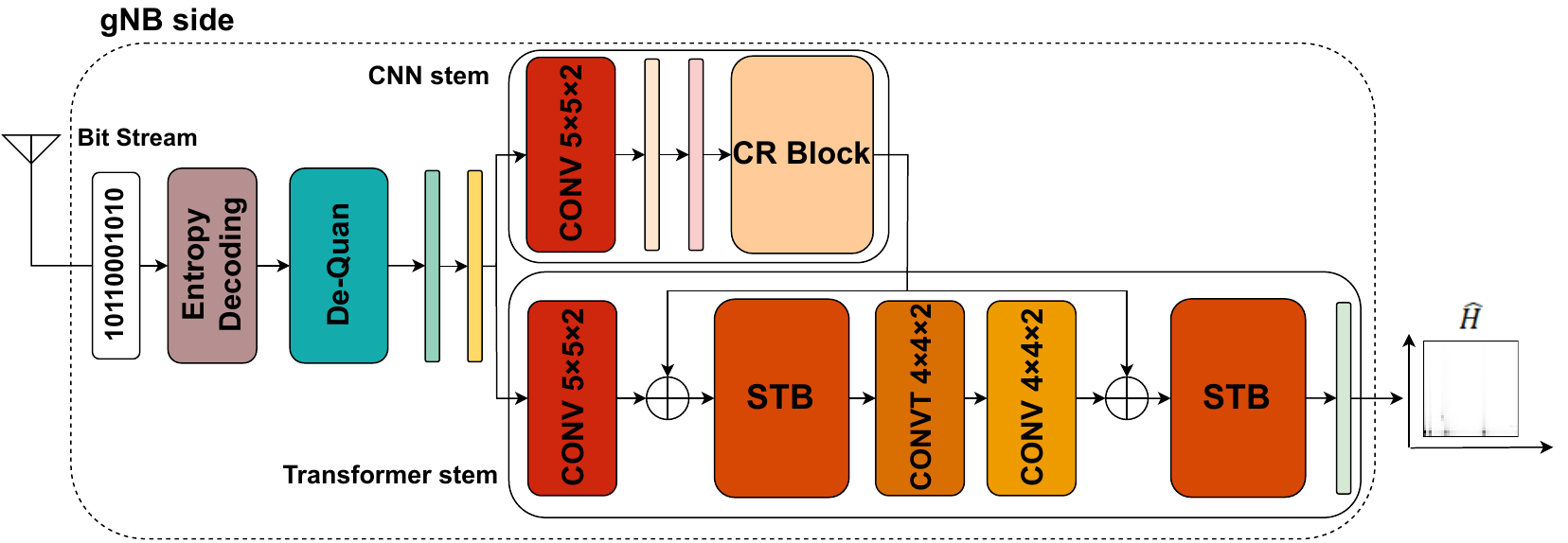}
    }
    \caption{STQENet architecture for CSI feedback. ``CONV'' and ``CONVT'' denote convolutional and transposed convolutional layers, respectively. ``STB'' refers to a spatially separable attention transformer block. ``CR Block'' represents the multi-resolution CNN module from CRNet (see Fig.~\ref{CRB}.)}
    \label{architecture}
    \vspace{-0.5cm}
\end{figure*}
\indent While many deep learning-based CSI compression methods enhance performance through architectural modifications, the CSI matrix is often treated as an unstructured data, thus neglecting the spatial correlations between antennas induced by the propagation environment. The attention mechanisms offer a more adaptive solution by explicitly modeling these correlations, enabling the network to focus on the most informative features. Notably, Attention-CSINet \cite{caiattention} pioneered the integration of attention mechanism into CNN-based CSI compression by introducing a module that generates a vector to describe the importance of each feature map, leading to superior performance. A two-layer transformer model, TransNet \cite{transnet}, demonstrated significant performance improvements but was ultimately deemed impractical due to its high computational complexity. To address this, CSIFormer \cite{csiformer} proposed a more efficient architecture employing locally grouped (windowed) self-attention, which lowered the computational burden, albeit with some reduction in performance. Furthermore,  \cite{lou2025attention} addresses the generalizability challenge of CSI compression by introducing AiANet, an attention-infused autoencoder that captures both global and local spatial features through self-attention mechanisms.\\
\indent In practical systems, CSI is fed back as bitstreams. Directly transmitting 32-bit floating-point codewords from the encoder would incur excessive overhead. To reduce this, codewords are quantized before feedback \cite{quan}, using uniform or non-uniform schemes to balance compression and accuracy. In \cite{chen2019novel}, the authors propose a bit-level CSI feedback framework that compresses downlink CSI using an encoder, followed by non-uniform quantization to produce a discrete bitstream. To mitigate quantization noise, a quantizer-dequantizer pair is introduced, along with an offset neural network that reconstructs the compressed CSI before decoding. This approach improves feedback accuracy but involves a complex three-stage training process, including pre-training the autoencoder, training the offset network while freezing other parameters, and a final fine-tuning phase. \cite{yin2025quantization} introduces a quantization module with bit allocation and propose a joint training method with an adaptive loss that balances quantization and reconstruction errors.\\
\indent Beyond quantization, entropy coding is an essential step for further reducing the bitrate by exploiting the statistical redundancy in quantized outputs. By assigning shorter codes to more probable symbols, entropy coding—such as Huffman or arithmetic coding—enables bit-level compression without information loss. While such frameworks mark a step toward practical CSI feedback, they typically treat quantization and entropy coding as separate post-processing steps, disconnected from the encoder-decoder training. This suboptimal separation can lead to performance degradation, as the encoder is not optimized to produce quantized representations that are both compact and entropy-efficient. \cite{ravula2021deep} introduces a simple AE-based CSI compression framework incorporating an entropy bottleneck to optimize quantization, entropy coding, and reconstruction quality jointly. Although effective, our method surpasses theirs in terms of reconstruction accuracy at lower bit rates.\\
\indent To address these limitations, we propose an end-to-end CSI compression framework, Spatially separable Transformer with Quantization and Entropy encoding Network (STQENet), that combines attention-based encoding, quantization, and entropy modeling to reduce feedback overhead in massive MIMO systems. The attention module captures inter-antenna correlations via spatially informative features, while quantization produces low-bit codewords for efficient transmission. A learned entropy model guides the compression by estimating the bit cost and introducing an entropy loss during training, enabling compact and accurate CSI representations. \\
\indent The main difference between our work and the state-of-the-art lies in the joint integration of attention-based encoding, quantization, and entropy modeling within a single end-to-end trainable framework. Unlike AiANet \cite{lou2025attention}, which focuses solely on attention mechanisms for spatial feature extraction without addressing quantization or entropy efficiency, our method incorporates both, enabling practical low-bitrate feedback. Compared to \cite{ravula2021deep}, which applies an entropy bottleneck to a standard autoencoder without attention modeling, our approach employs a spatially separable transformer to better capture antenna-domain correlations. Experiments show that our method achieves superior rate-distortion performance compared to state-of-the-art transformer-based approaches, with lower bitrate and complexity.\\
\indent The remainder of the paper is structured as follows. Section II introduces the system model. Section III describes the proposed architecture in detail. Section IV presents the experimental results along with a comprehensive discussion. Finally, Section V concludes the paper.

\section{System Model}
We consider an FDD system where the gNB is equipped with $N_t \gg 1$ transmit antennas and a single
receiver antenna at a User Equipment (UE). The system operates in Orthogonal Frequency Division Multiplexing (OFDM) with $\tilde{N}_c$ subcarriers. The received signal at the $n$-th subcarrier is provided as follows:
\begin{align}
    y_n = \tilde{h}_n^H \mathbf{v}_n x_n + z_n,
\end{align}
where $\tilde{h}_n \in \mathbb{C}^{N_t \times 1}$, $\mathbf{v}_n \in \mathbb{C}^{N_t \times 1}$, $x_n \in \mathbb{C}$, $z_n \in \mathbb{C}$ denote the channel vector, precoding vector, data-bearing symbol, and additive noise of the $n$-th subcarrier, respectively. The CSI matrix in the spatial-frequency domain is given by $\tilde{\mathbf{H}} = [\tilde{\mathbf{h}}_1, \tilde{\mathbf{h}}_2, \ldots, \tilde{\mathbf{h}}_{\tilde{N}_c}]^H\in \mathbb{C}^{\tilde{N}_c \times N_t}$. The number of feedback parameters is $2\tilde{N}_c N_t$. To reduce feedback overhead, we adopt the compression approach from \cite{csinet}, which sparsifies $\tilde{\mathbf{H}}$
in the angular-delay domain using a $2$D Discrete Fourier Transform (DFT):
\begin{align}
    \bar{\mathbf{H}} = \mathbf{F_d} \tilde{\mathbf{H}}\mathbf{{F}_a^H}, 
\end{align}
where $\mathbf{F_d}$ and $\mathbf{F_a}$ are $2$D DFT matrices of dimensions $\tilde{N}_c \times \tilde{N}_c$ and $N_t \times N_t$, respectively. In the delay domain, the time delay between multipath arrivals is typically confined to a limited range, resulting in most of the significant energy being concentrated in the first \( N_c \leq \tilde{N}_c \) rows of the channel matrix $\bar{\mathbf{H}}$. Assuming $\bar{\mathbf{H}}$ is complex-valued and exhibits similar sparsity, we retain its first $N_c$ rows and then concatenate the real and imaginary parts to construct a real-valued matrix $\mathbf{H}\in \mathbb{R}^{2N_c \times N_t}$. With the sparsified channel matrix $\mathbf{H}$ obtained, it is processed through the encoder-decoder architecture depicted in Fig. \ref{architecture} In this stage, $\mathbf{H} $ is compressed into a one-dimensional vector of size \( M \times 1 \). The compression ratio is defined as $\gamma = \frac{M}{2N_cN_t}$. To reduce the bitrate and prepare for practical transmission, the continuous-valued latent vectors are passed through a quantization module. This step maps the continuous values to discrete symbols. After quantization, the discrete symbols are entropy encoded using Huffman coding. Huffman encoding assigns shorter codewords to more frequent symbols, minimizing the average code length. This lossless compression step ensures that the quantized data is stored or transmitted with minimal redundancy.

This compressed representation is then transmitted from the UE to the gNB over the uplink. Upon reception, Huffman decoding and de-quantization are first applied to recover the compressed latent features, which are then passed through the decoder to reconstruct the channel matrix, denoted as $\hat{\mathbf{H}}$. The encoding and decoding processes are defined as follows:
\begin{align}
    \nonumber
    \mathbf{s} = f_e(\mathbf{H}), \quad \hat{\mathbf{H}} = f_d(\mathbf{s}),
\end{align}
where $f_e$ and $f_d$ represent the functions of the encoder and decoder, respectively. Here, $\mathbf{s}$ denotes the compressed codeword, and $\hat{\mathbf{H}}$ is the reconstructed channel matrix estimated by the model.
\section{STQENet architecture}
\begin{figure*}[t]
  \centering

  \subfloat[LSA and GSA blocks\label{lsagsa}]{
    \includegraphics[width=0.4\textwidth]{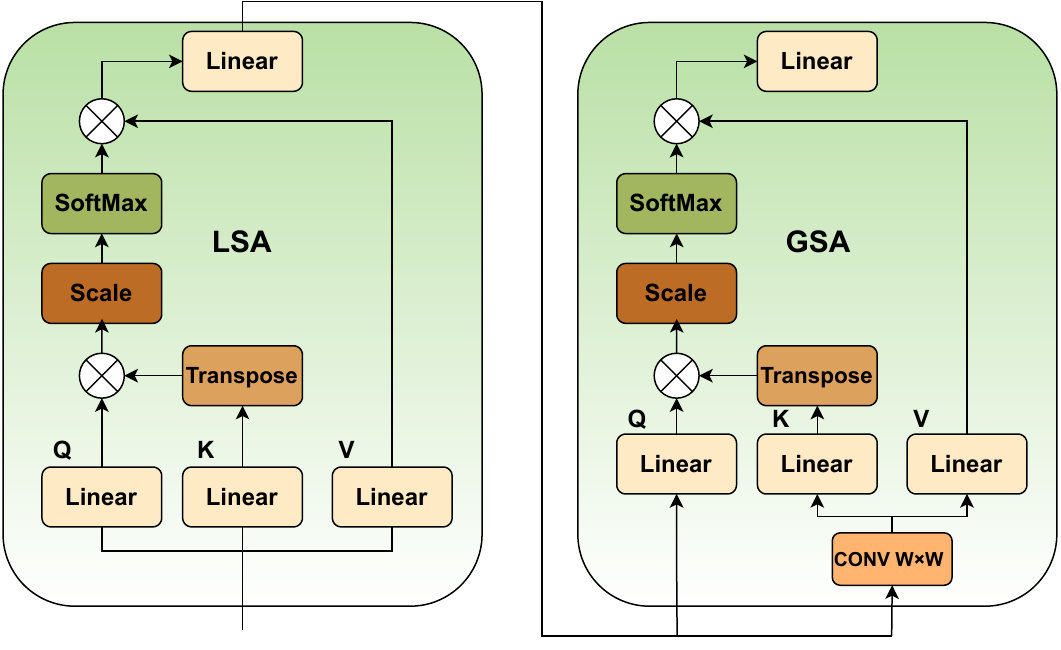}
  }
  \hfill
  \subfloat[STB block\label{stb}]{
    \includegraphics[width=0.2\textwidth]{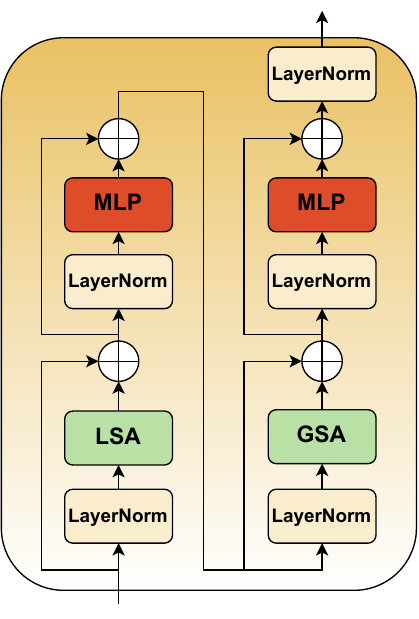}
  }
  \hfill
  \subfloat[CR block\label{CRB}]{
    \includegraphics[width=0.3\textwidth]{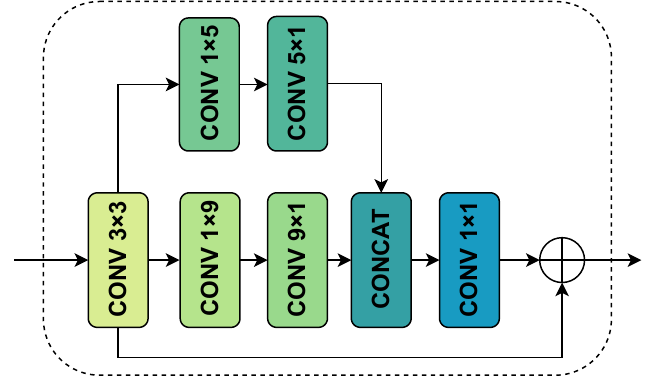}
  }

  \caption{Detailed blocks of the STQENet architecture, adopted from the STNet architecture~\cite{STNet}.}
  \label{blocks}
\end{figure*}

We implement the CSI encoder and decoder within STQENet, as shown in Fig. \ref{architecture}, drawing inspiration from the STNet architecture \cite{STNet}. Leveraging the attention mechanism, STNet achieves high performance in CSI feedback tasks. It effectively reduces the amount of feedback data while aggregating spatial-frequency domain CSI features. Moreover, its architecture captures long-range dependencies and exploits inter-antenna correlations within the channel matrix, enhancing the accuracy of the reconstructed CSI. \cite{STNet} presents a detailed description of the STNet architecture. In this Section we explain each block in the architecture.
\subsection{Spatially Separable Attention Transformer Block (STB)}
The STB incorporates a hybrid attention mechanism consisting of three main stages: Locally-grouped Self-Attention (LSA), Global Sub-sampled Attention (GSA), and Multi-Layer Perceptron (MLP).
\subsubsection{LSA}
In the LSA stage, the input channel matrix (of size $L \times L$) is divided into non-overlapping windows of size $W \times W$, where $W = L/m$ and $m$ is the number of partitions along each spatial dimension. Attention is computed independently within each window, reducing the complexity from $\mathcal{O}({L^4}d)$ to $\mathcal{O}(\frac{L^4}{m^4}d)$,  where $d$ is the feature dimension. However, because attention is local and restricted to each window, global dependencies across windows are lost.
\subsubsection{GSA}
To address this limitation without reverting to full global attention, the GSA mechanism is introduced. The output from LSA is passed through a CNN layer with stride $W$, effectively summarizing each window into a single spatial token and producing a feature map of size $m \times m$. This map serves as the keys and values for another layer of attention, while the LSA output continues to serve as the queries. The GSA thus reintroduces global context in a computationally efficient manner, with a complexity of $\mathcal{O}(m^2L^2d)$. The total complexity of the full attention mechanism becomes:
\begin{align}
    \mathcal{O}\left(\frac{L^4}{m^4}d + m^2L^2d\right)
\end{align}
\subsubsection{MLP block}
MLPs are essential for feature transformation and nonlinear mixing across channels (not across spatial locations, which is done by attention layers). This block has a linear layer followed
by a Gaussian Error Linear Unit (GELU) non-linearity and another linear layer.
\subsection{CR Block}
Beyond the attention block, the overall decoder architecture consists of two parallel stems: one transformer-based and the other convolution-based. This dual-branch design is intended to leverage the strengths of both components—while the transformer branch effectively captures global dependencies, the convolutional branch focuses on modeling local spatial features and enhances generalization through weight sharing and spatial invariance. As illustrated in Fig.~\ref{CRB}, the convolutional branch integrates multi-kernel convolutions to extract spatial features at multiple scales, further enriching the feature representation.
\subsection{Quantization and Entropy}
\begin{figure*}[h]
    \centering

    \subfloat[indoor dataset\label{snr1}]{
        \includegraphics[width=0.48\textwidth]{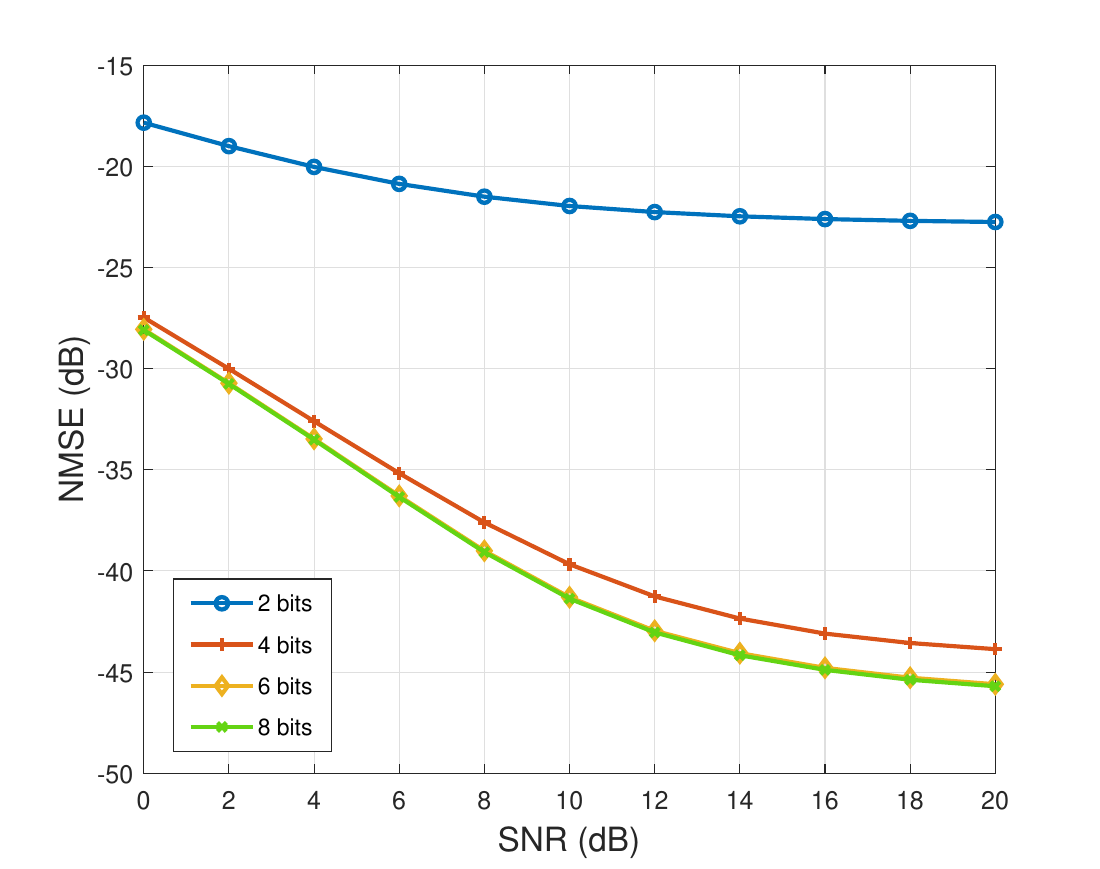}
    }
    \hfill
    \subfloat[ outdoor dataset\label{snr2}]{
        \includegraphics[width=0.48\textwidth]{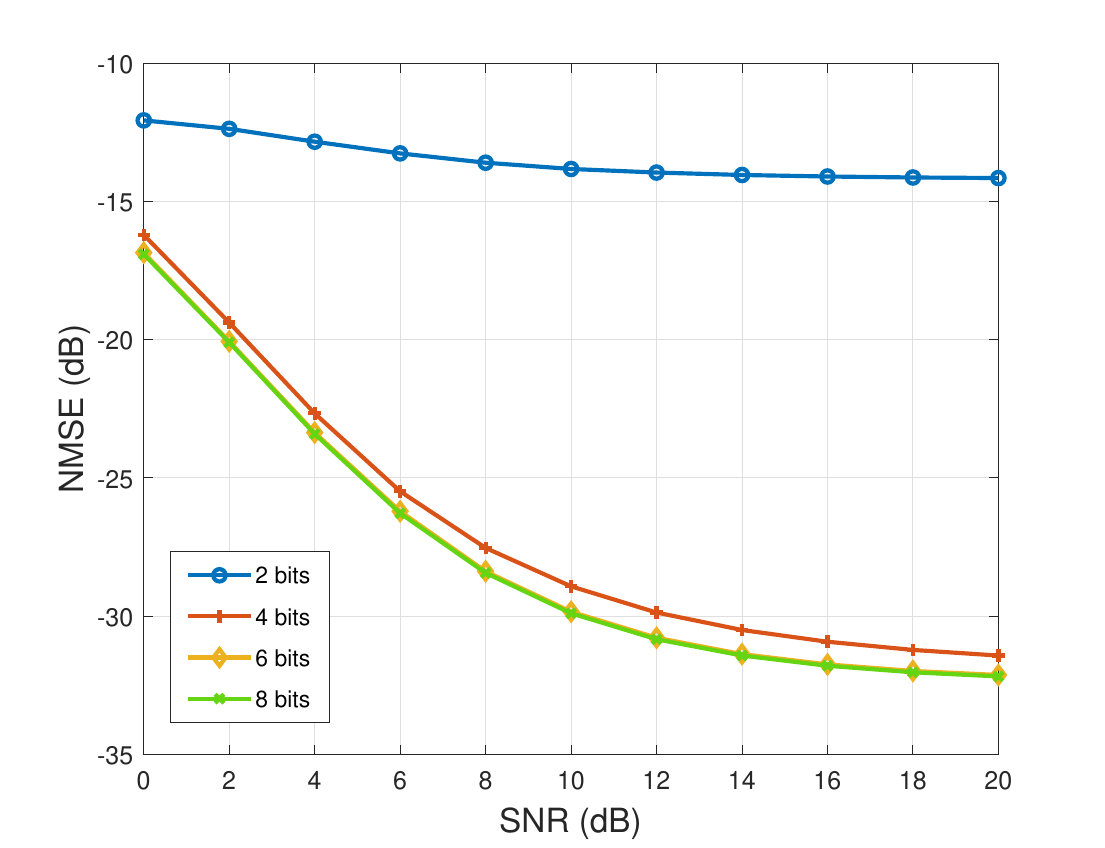}
    }

    \caption{NMSE vs. SNR at different quantization bit-rates.}
    \label{snr}
      \vspace{-0.5cm}
\end{figure*}

To enable end-to-end differentiable training while preserving the benefits of entropy coding, we adopt a learned probabilistic quantization model based on $\mu$-law companding \cite{jang2019deep}. To address the non-differentiability of the quantization function, which poses a challenge for training neural networks via backpropagation, we employ the Straight-Through Estimator (STE) technique. In this approach, the true gradient is replaced with a constant during backpropagation, effectively enabling gradients to pass through the quantization step \cite{ste}. For a normalized output of the encoder, the $\mu$-law compression function is defined as:
\begin{align}
    f(s) =  \operatorname{sgn}(s) \frac{\ln(1 + \mu |s|)}{\ln(1 + \mu)}.
\end{align}
\indent Entropy is computed as the expected negative log-likelihood of the quantized latent variables given $\hat{s}$:
\begin{align} \label{entropy}
    R = \mathbb{E}_{H \sim p_H} \left[ -\log_2 \hat{p}_s(\hat{s}) \right],
\end{align}
where $p_H$ is the true distribution of CSI tensors and $\hat{p}_s$ models the probability of each quantized value. In our implementation, we adopt a factorized prior for the quantized latent variables. Specifically, $\hat{p}_s$ is modeled using an empirical histogram of symbol frequencies, and entropy coding is performed with Huffman coding.
\section{Experiments and Numerical Results}
\begin{figure*}[h]
    \centering

    \subfloat[ indoor dataset\label{BPP_indoor}]{
        \includegraphics[width=0.48\textwidth]{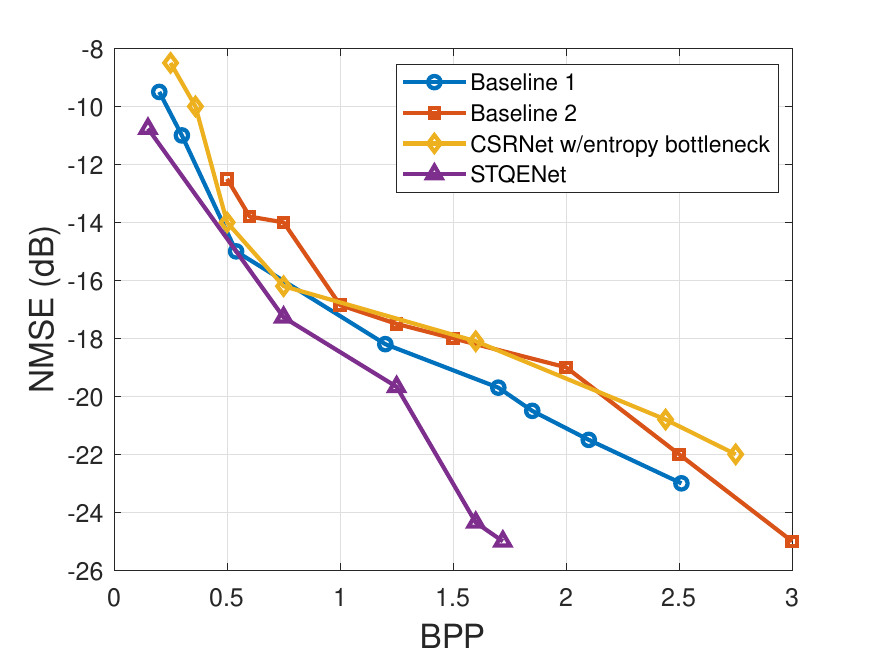}
    }
    \hfill
    \subfloat[ outdoor dataset\label{BPP_outdoor}]{
        \includegraphics[width=0.48\textwidth]{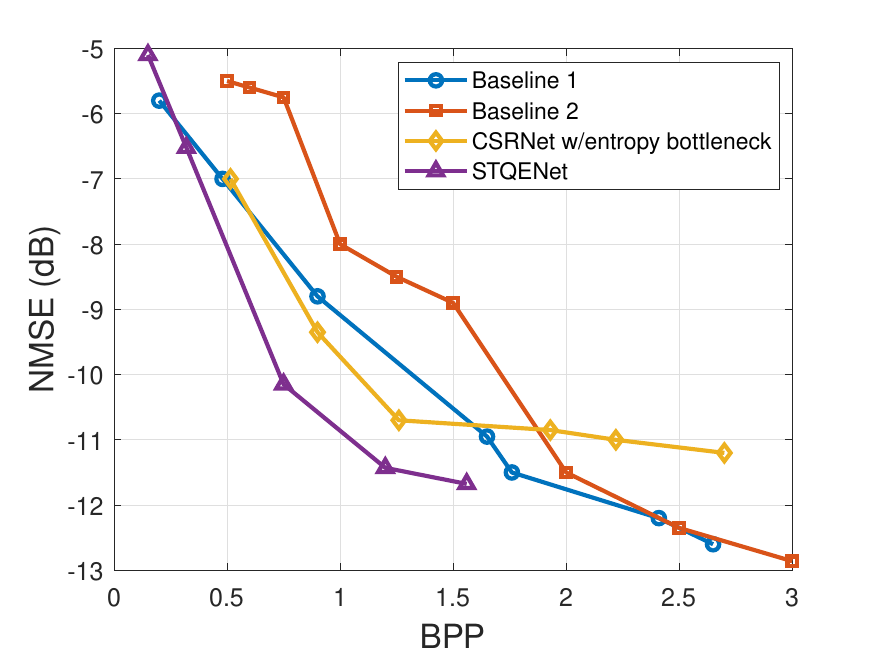}
    }

    \caption{NMSE vs BPP for different methods.}
    \label{BPP}
      \vspace{-0.5cm}
\end{figure*}

We examine a system configuration featuring $32$ antennas at the gNB and a single antenna at the UE. For performance evaluation, we utilize the COST2100 dataset~\cite{dataset}, focusing on two specific scenarios: an indoor picocellular environment operating at $5.3\,\text{GHz}$, and an outdoor rural setting at $300\,\text{MHz}$. The number of subcarriers is set to $N_c = 32$, with a window size $W = 8$, and the multi-head attention mechanism employs $P = 4$ heads. The dataset is divided into $100{,}000$ samples for training, $30,000$ for validation, and $20{,}000$ for testing. We use a batch size of $200$ and train the model for $1000$ epochs. The optimization process is carried out using the Adam optimizer with a learning rate of $0.001$, and the Mean Squared Error (MSE) and entropy loss are used as the loss function, as follows;
\begin{align}
    \mathcal{L}(\theta_e, \theta_d, \phi) = \frac{1}{B} \sum_{i=1}^{B} \| \mathbf{H}_i - \hat{\mathbf{H}}_i \|^2 + \lambda R,
\end{align}
where $\mathbf{H}$ is the input channel matrix, $\hat{\mathbf{H}}$ is the reconstructed channel matrix, and $B$ is the batch size. $\theta_e$, $\theta_d$, and $\phi$ are the parameters of the encoder, decoder, and the entropy bottleneck, respectively. $\lambda$ is the regularization parameter that decides the rate–distortion tradeoff, which is set to $10^{-3}$. We adopt the Normalized Mean Squared Error (NMSE) as the evaluation metric defined as follows:
\begin{align}\label{nmse}
   \text{NMSE} = \mathbb{E} \left\{ \frac{\| \mathbf{H} - \hat{\mathbf{H}} \|^2}{\| \mathbf{H} \|^2} \right\}.
\end{align}
\indent We apply \(10 \log_{10}(\cdot)\) to the expectation in Eq.~\ref{nmse} and report the NMSE in decibels (dB).  
Entropy is expressed in Bits Per Pixel (BPP), calculated by dividing the entropy value from Eq.~\ref{entropy} by $2 \times 32^2$ ($2$ refers to the real and imaginary part of the CSI data),  
which represents the spatial dimension of the CSI tensors. For the results in Fig. \ref{BPP}, the BPP is computed as:
\begin{align}
    \text{BPP} = \frac{k \times N}{2 \times 32^2},
\end{align}
where $N$ and $k$ are the number of quantization bits, and encoder output dimension. The source code for the simulation of our approach is available in \cite{code}.

\subsection{Result Analysis}
We evaluate the reconstruction quality using the NMSE metric across a range of Signal-to-Noise Ratio (SNR) values from 0 dB to 20 dB. The evaluation is performed on both indoor and outdoor datasets under different quantization bit-rates (2, 4, 6, and 8 bits), as shown in Fig. \ref{snr}. \\\
\indent We observe in Fig. \ref{snr1} that as the quantization bit-rate increases, the NMSE improves consistently across all SNR values.  For example, the 8-bit quantization significantly outperforms the 2-bit version, especially at higher SNRs, confirming that finer quantization granularity leads to better channel reconstruction. For a fixed bit-rate, the NMSE decreases as SNR increases. This indicates that the model benefits from better signal quality, leading to more accurate CSI recovery. The curve corresponding to 2-bit quantization exhibits a performance floor, especially at higher SNRs. This suggests that coarse quantization becomes the dominant source of distortion, limiting the benefits from improved SNR. The trends observed in Fig. \ref{snr2} for the outdoor dataset align closely with the indoor scenario, with some noticeable differences: The NMSE values in the outdoor dataset are generally higher than those of the indoor dataset at the same bit-rate and SNR. This is due to the more complex and variable propagation conditions in outdoor environments, which increases the difficulty of accurate CSI recovery. The performance gap between different quantization levels is more pronounced in the outdoor dataset. For example, the improvement from 2 bits to 8 bits is larger in the outdoor case compared to the indoor case, highlighting the importance of sufficient quantization resolution in challenging environments.\\
\indent Figure \ref{BPP} shows the NMSE performance versus BPP for various methods evaluated on both indoor and outdoor datasets. The compared models include Baseline 1 \cite{ravula2021deep}, Baseline 2 \cite{chen2019novel}, CSRNet \cite{crnet}, and the proposed STQENet architecture. To ensure a fair comparison, we integrated an entropy bottleneck into Baseline 2 and CSRNet, enabling them to function as end-to-end frameworks similar to our approach. For indoor dataset in Fig. \ref{BPP_indoor}, STQENet consistently outperforms all baseline methods across most BPP values, especially in the mid-to-low rate region (0.5–1.5 BPP). This indicates that our method, achieves superior compression efficiency without sacrificing reconstruction quality. STQENet outperforms Baseline 1, which uses a simple autoencoder with different quantization but the same entropy model, by an average of $17.72\%$, demonstrating the benefits of using attention mechanisms in the autoencoder for better feature representation. Baseline 2, focusing solely on quantization without an entropy model, performs significantly worse, which STQENet outperforms it by $27.48 \%$. This confirms the critical role of entropy modeling in achieving efficient compression. CSRNet with an entropy bottleneck shows competitive performance but still lags behind STQENet, which beats it by an average of $19.32\%$. This suggests that while CSRNet benefits from added entropy modeling, its underlying autoencoder may not be as expressive or adaptive as the attention-based design in STQENet. In the outdoor dataset, shown in Fig. \ref{BPP_outdoor}, STQENet once again delivers top-tier performance at low BPP levels (around ~0.75), demonstrating strong robustness in handling more challenging compression scenarios. It exhibits a significant NMSE reduction as BPP rises, highlighting its effectiveness in environments with limited bandwidth. STQENet delivers moderate to substantial average improvements across benchmarks: $8.79\%$ over Baseline 1, $40.11\%$ over Baseline 2, and $7.68\%$ over CSRNet with entropy bottleneck. The largest gain is observed against Baseline 2, highlighting the significant advantage of integrating entropy modeling with learned attention-based compression.

\section{Conclusion}
In this paper, we introduced STQENet, an end-to-end CSI compression framework designed to address the feedback bottleneck in massive MIMO systems. By leveraging a Spatial Correlation-Guided Attention Mechanism alongside quantization-aware and entropy-aware training, our approach effectively captures antenna dependencies and produces compact, efficiently coded representations. Experimental results across diverse datasets demonstrate that STQENet consistently outperforms existing compression methods. Our analysis reveals that performance improves steadily with finer quantization and higher SNR, while coarse quantization introduces distortion floors, especially in high-SNR or complex propagation scenarios. Compared to baselines with similar entropy modeling, STQENet achieves superior rate-distortion trade-offs due to its attention-based architecture and learned coding pipeline. These findings underscore the importance of jointly optimizing neural compression models with practical coding constraints in mind. Our work offers a promising step toward scalable and deployable CSI feedback solutions for next-generation wireless networks.

\section*{Acknowledgment}
This work is funded by the Dutch National Growth Fund project “6G Future Network Services (FNS)” and Eindhoven Hendrik Casimir Institute (EHCI), Netherlands; and supported in part by the Department of Science, Innovation and Technology, U.K., under Grant Yorkshire Open-RAN (YO-RAN).
\bibliographystyle{IEEEtran}
\bibliography{IEEEabrv,Bibliography}

\vspace{12pt}

\end{document}